\documentclass[runningheads,a4paper]{llncs}

\usepackage{amssymb}
\usepackage{graphicx}

\usepackage{placeins}

\usepackage{url}
\urldef{\mailsa}\path|daniel.ritter@sap.com|    
\newcommand{\keywords}[1]{\par\addvspace\baselineskip
\noindent\keywordname\enspace\ignorespaces#1}

\usepackage{tabularx}

\usepackage{makeidx}
\makeindex

\newcommand{\eg}{e.\,g.}
\newcommand{\ie}{i.\,e.}

\begin{document}

\mainmatter  

\title{Using the Business Process Model and Notation for Modeling Enterprise Integration Patterns}

\titlerunning{Using the BPMN for Modeling Enterprise Integration Patterns}

%
%
\author{Daniel Ritter}
%

\institute{HANA Platform, SAP AG\\
Dietmar-Hopp-Allee 16, 69190 Walldorf, Germany\\
\mailsa}

%
%

\toctitle{Lecture Notes in Computer Science}
\tocauthor{Authors' Instructions}
\maketitle

\begin{abstract}
\emph{Enterprise Integration Patterns} (EIP) are a collection of widely used stencils for integrating enterprise applications and business processes. These patterns represent a ``de-facto" standard reference for design decisions when integrating enterprise applications.

For each of these patterns we present the integration semantics (model) and the conceptual translation (syntax) to the \emph{Business Process Model and Notation} (BPMN), which is a ``de-facto" standard for modeling business process semantics and their runtime behavior.
\end{abstract}

\section{Introduction}
In the last decades, companies have developed a multitude of software applications. The interest of reusing them is not only a matter of convenience, but also of costs. Reuse of applications often occurs in connection with third parties--another department, or even another company. To allow applications from different domains--namely different data types and interfaces--to communicate, all parties have to integrate with a communication solution.

In their best practices book Enterprise Integration Patterns (EIP) \cite{Hohpe:2003:EIP:940308} have collected a widely used and accepted collection of integration patterns allowing for easier implementation of a communication infrastructure. They identify four possible solutions for solving communication: file transfer, shared database, remote procedure invocation, and messaging. The patterns presented in the book are typical concepts used when implementing a messaging system and have proved to be useful during implementation. These concepts help to cope with the asynchronous nature of message exchange and the facts, that ``networks are unreliable", ``networks are slow", ``any two applications are different", and ``change is inevitable". On the other hand, the modular nature of patterns allows them to be used efficiently in new implementations. Subsequently, we present most patterns described by \cite{Hohpe:2003:EIP:940308} plus well-known additional patterns (\eg, from \cite{DBLP:phd/de/Scheibler2010} and \cite{apacheCamel13}) and give a realization as Business Process Model and Notation (BPMN) \cite{BPMN20}. Thus we have business process building blocks allowing us to model a messaging system in BPMN that is based on a design given by EIPs.

\begin{figure}
\centering
\includegraphics[width=1.0\columnwidth]{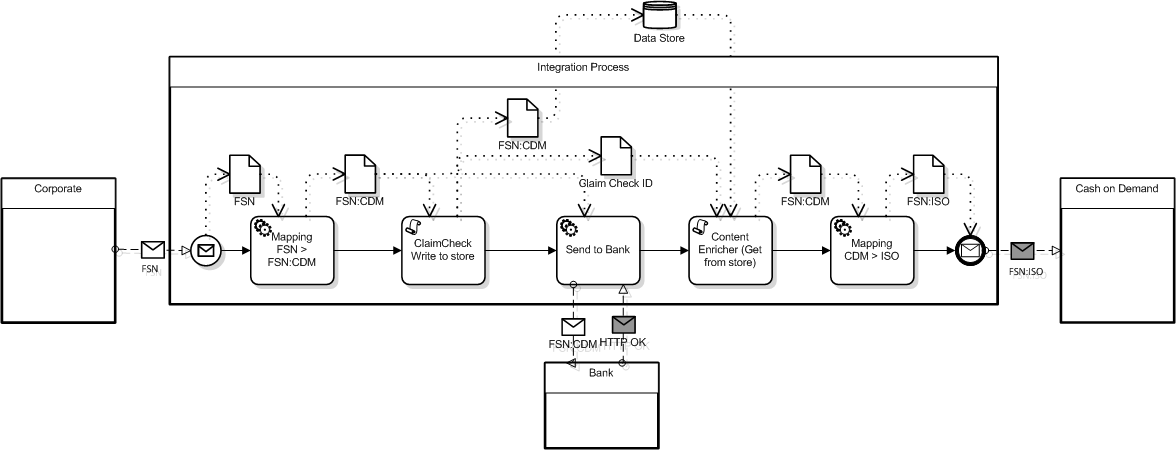}
\caption{Business Monitoring: Messages sent from Corporate to Bank are routed to SAP Cloud for Cash for business monitoring purpose.}
\label{fig:cod}
\end{figure}

For instance, Figure \ref{fig:cod} shows an integration scenario of a corporate with its bank and business monitoring via SAP Cloud for Cash, syntactically expressed in BPMN according to the definition proposed by this paper. The incoming message is of type ``FSN" (short for Financial Services Network\footnote{http://scn.sap.com/docs/DOC-40696}), which has to be translated to its canonical data model incarnation ``FSN:CDM" using a \emph{Message Translator}\index{Message Translator (85)} pattern. Through an adapted \emph{Claim Check}\index{Claim Check (346)} pattern, the message is stored for later use and handed over to the \emph{External Service}\index{External Service}\index{External Service} pattern as request to the bank (no further translation required). On successful execution, the original message is restored from the claim check, translated to an ISO format ``FSN-ISO" and send to the SAP Cash on Demand application\footnote{http://global.sap.com/germany/solutions/technology/cloud/business-by-design/highlights/index.epx} (ODC).

The problem that we address in this paper is the lack of standardized syntactic and semantic formalization of the common integration patterns \cite{Hohpe:2003:EIP:940308} and recent extensions (\eg, \emph{Join Router}\index{Join Router}, \emph{External Service}\index{External Service} from \cite{DBLP:phd/de/Scheibler2010}, \emph{Synch-Asynch/ Asynch-Synch}\index{Synch-Asynch Bridge}\index{Asynch-Synch Bridge} bridges from \cite{volkersbuch}, and \emph{Multicast}\index{Multicast} from \cite{apacheCamel13}). Our goal is to provide means for an implementation independent description (\ie, syntax) and semantics (\ie, model) of a messaging system. That means, only the modeling is considered and the concrete runtime implementations are neglected. Our BPMN realization provides a blue print for implementation that is proven to translate to existing messaging systems like Apache Camel \cite{apacheCamel13}, SAP HANA Cloud Integration\footnote{http://help.sap.com/cloudintegration}, and SAP Process Integration\footnote{http://scn.sap.com/community/pi-and-soa-middleware}.

The remainder of this paper is as follows: Section \ref{sec:relatedwork} discusses and manifests the contribution of the paper by putting it into context to related work. Then, integration semantics and the used subset of BPMN is defined in the context of the EIPs, which leads to the definition of the syntax and model in Section \ref{sec:eips}.
Section \ref{sec:conclusion} concludes the discussion with a short summary.




\section{Related Work}\label{sec:relatedwork}
The patterns described by \cite{Hohpe:2003:EIP:940308,DBLP:phd/de/Scheibler2010,volkersbuch,apacheCamel13} are not building blocks of a modeling language, however, they describe typical concepts in designing a messaging system; thus they are an informal specification language. For that, there are elaborated modeling techniques like the \emph{Business Process Model and Notation} (BPMN) \cite{BPMN20} or the \emph{Workﬂow Patterns} defined by \cite{DBLP:journals/dpd/AalstHKB03}. Enterprise Integration Patterns complement these notations by a set of typical designs found in a messaging infrastructure.

Our approach stresses on the control flow, data flow and modeling capabilities of BPMN as well as its execution semantics. Recent work on ``Data in Business Processes" \cite{DBLP:journals/emisa/MeyerSW11} shows that besides \emph{Configuration-based Release Processes} (COREPRO) \cite{DBLP:conf/otm/MullerRH07,DBLP:phd/de/Muller2009,DBLP:conf/bpm/MullerRH06}, which mainly deals with data-driven process modeling and (business) object status management, and UML activity diagrams, BPMN achieves the highest coverage in the categories relevant for our approach. Compared to BPMN and apart from the topic of ``object state" representation, neither \emph{Workflow Nets} \cite{DBLP:journals/jcsc/Aalst98} nor petri nets do support data modeling at all \cite{DBLP:journals/emisa/MeyerSW11}. Based on that work, BPMN was further evaluated with respect to data dependencies within BPMN processes \cite{DBLP:conf/bpm/MeyerPFW13,DBLP:conf/er/MeyerW12}, however, not towards control and data flow as in our approach.

Our work combines this foundation with efforts on executable integrations patterns, their configuration and mapping to the \emph{Web Services Business Process Execution Language}\footnote{https://www.oasis-open.org/committees/tc\_home.php?wg\_abbrev=wsbpel} (WSBPEL) proposed by \cite{DBLP:phd/de/Scheibler2010} and leverages the work of \cite{tuprints2751,volkersbuch} that started to map the EIPs to the BPMN syntax and some semantics by example. In this document, we provide a systematic continuation of \cite{tuprints2751,volkersbuch} by defining a comprehensive syntax and model for widely-used patterns.

Since BPMN can be translated to \emph{Petri Nets} \cite{petri1962,DBLP:conf/vveis/RaedtsPUWGS07}, our approach can leverage the body of work from the BPMN verification domain. In particular, the area of verification of middleware designs \cite{DBLP:conf/caise/FahlandG13} translated the EIPs from \cite{Hohpe:2003:EIP:940308} into \emph{Colored Petri Nets}. Since BPMN is closely related to petri nets, the syntactical translation is comparable to our approach. However, the concrete syntactical mapping of integration patterns and their corresponding model (semantics) that leverages the BPMN syntax with a partially changed execution model is novel.


\section{Enterprise Integration Patterns} \label{sec:eips}
Before presenting the patterns we shortly discuss the BPMN syntax that is used to describe EIPs and the (process) model we use for the integration semantics. The model might slightly differ from the original BPMN runtime semantics.

The main syntactical artifacts in BPMN conduct process steps, sequences and the representation of messages that are exchanged between processes during runtime. Subsequently, the BPMN elements relevant to this work are introduced. They are mainly taken from the \emph{BPMN Collaboration Diagram} \cite{BPMN20}.

\begin{figure}
\centering
\includegraphics[width=0.8\columnwidth]{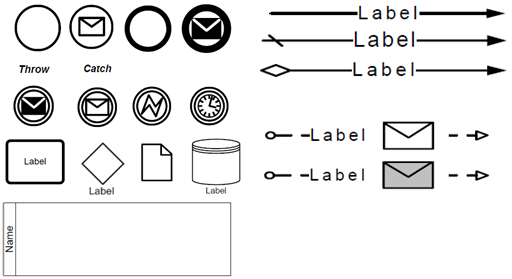}
\caption{Used BPMN Syntax.}
\label{fig:bpmn_elements}
\end{figure}

\noindent
\textbf{Process flow control}
The flow in a BPMN graph is controlled by flow objects that initiate, route and terminate the flow:
\begin{itemize}
\item Events: control the process flow. We consider start and end events. Start events initiate the process flow and thus only have outbound edges in the graph. Figure \ref{fig:bpmn_elements} shows a start event without defined event (first row, left) and one with a ``message-arrives" semantic, which triggers the event (first row, second from left). Similarly, the end event terminates the flow and thus terminates a process or part of a process. The standard end event without defined result (first row, third from left) can also appear with message symbol with the semantic of sending a message to a participant before terminating the process (first row, forth from left). In addition, throwing/catching intermediate events are used to express message events (second row, two icons on the left), errors (second row, middle), and timers (second row, right).
\item Activities/Tasks: are process steps that have to be finished, before a flow can proceed. Depending on the type of activity, a process step can be executed multiple times. The label specifies the operation that is executed by the process step (third row, left). If the activity is atomic (\ie, a task), the task types like \emph{Send Task}, \emph{Receive Task}, \emph{Script Task} can be specified (not shown).
\item Gateways: handle multiple process flows, where they work as logical cases that support simple logical operations like AND, OR, and more complex nested cases within flows (\eg, parallel processing, multiplexers). In this document, we mainly work with exclusive (third row, second from left), inclusive and parallel gateways where the label specifies the semantics of the case.
\end{itemize}

\noindent
\textbf{Connections}
The connecting objects are the edges in the graph that differ by the transferred objects:
\begin{itemize}
\item Sequence Flows: sequences of process steps. The non-default, standard sequence flow has an unspecified ``conditionExpression" (right, top). The default sequence flow again has an unspecified ``conditionExpression" (right, second to top). A specified condition expression with non-default semantics is shown in Figure \ref{fig:bpmn_elements} (right, third to top).
\item Message Flow: message exchange (\eg, process status information, error messages, data) between participants or participants and process elements. The initiating message flow is shown on the left, and the non-initiating message flow on the right.
\end{itemize}

\noindent
\textbf{Pools}
\begin{itemize}
\item Pool/ Participant:  a pool is a general graphical, grouping element, which represents a participant. It also acts as a graphical container for partitioning a set of Activities from other pools, usually in the context of B2B situations. A pool may have internal details, in the form of the process that will be executed. A pool that has no internal details is called ``black-box" pool (Figure \ref{fig:bpmn_elements}, fourth row).
\end{itemize}

\noindent
\textbf{Additional Artifacts}
\begin{itemize}
\item Data Objects: represent provided or generated Data. Data objects are associated (cf. association) to another BPMN element and do not exist independently. Figure \ref{fig:bpmn_elements} shows data objects (third row, third from left). The data associations are not shown.
\item Data Stores: represent a local or central storage of data (\eg, messages). Stores are associated to BPMN Flow Elements (\eg, Data Objects, Events, Activities, Gateways). The data store is shown in Figure \ref{fig:bpmn_elements} (third row, fourth from left).
\end{itemize}

Besides the syntax (\ie, graphical notation), BPMN specifies a (process) model (\ie, semantics), called execution semantics. The execution semantics define the meaning of the syntax and specify the behavior during execution. In this work, we remain conformant to the syntax, but define a slightly different process model for the execution semantics (basically a subset of the BPMN semantics with a stronger focus on the data flow).




\begin{definition}[Process model]
A \emph{process model} $M = (N,SF,DO,DF)$ consists of a finite non-empty set $N \subseteq A \cup G \cup E$ of nodes being activities A of types ServiceTask, ScriptTask and MessageTask, gateways G of types ExclusiveGateway and ParallelGateway, and events E of types StartEvent, EndEvent and Intermediate Event, where A, G, E are pairwise disjunct.

The finite non-empty set of SequenceFlow relations $SF \subseteq (N \setminus {EndEvent}) \times (N \setminus {StartEvent})$ represents the control flow. The finite, non-empty set of data objects $DO$ represents data associated to $N$ and $DF \subseteq (N \cup DO) \times (DO \cup N)$ is the data flow relation.

For the Process and Sub-Process instantiation, a Message \emph{Start Event} or \emph{Receiving Message Task} is required comparable to a \texttt{constructor}. The instances can be terminated by \emph{(Message) End Events}, the \texttt{destructor}. An already instantiated process can be re-invoked using the BPMN correlation mechanism, similar to a \texttt{factory pattern}.
\end{definition} 




In other words, a process is initiated by a start event, \ie, a message that contains data according to a specified format (\eg, XML Schema). Then a sequence flow is fired that moves the control to the next flow element in the process (\eg, acticity, gateway, end event). The data flow is handled by associated data objects from one element to the next one. The process ends through the invocation of a message end event that fires the outgoing message before the process context stops.

\subsection{Integration Semantics}\label{sec:integrationsemantics}
The integration patterns that are mapped to BPMN subsequently can be summarized by their integration semantics like (i) cardinality, (ii) message and meta-data creation, (iii) quality of service/ transactional behavior, (iv) logic, (v) lineage, and (vi) processing types.

The \emph{cardinality} (i) can be determined on a message and a channel level. On the message level, the cardinality specifies the number of incoming and outgoing message instances of a pattern. For instance, if the original message is routed through an \emph{Activity} representing a pattern, it has a cardinality of 1:1 and if a pattern splits the original message into $n$ outgoing messages, then the cardinality is 1:n. Similarily, the definition of cardinality on a message channel level describes the incoming and outgoing channels (edges) of a pattern. Note that the \emph{Splitter Pattern}\index{Splitter (259)} has a message cardinality of 1:n, but a channel cardinatlity of 1:1 (\ie, 1:n if and only if the multiple output formats are defined).

\emph{Message Creation} (ii) is a characteristic that defines the behavior of a pattern when processing the original message. If and only if the message cardinality is 1:1 and the identifier of the outgoing is the same as the incoming message instance (not necessarily the content or format), a pattern is \emph{non-message generating}. Otherwise, the pattern is \emph{message generating}. The message's meta-data is either represented by a \emph{Message.ItemDefinition} assigned to the incoming and outgoing \emph{MessageFlow} elements or a \emph{DataObject.ItemDefinition.structureRef} within the \emph{Process} and its \emph{Sub-Process} definitions.

The Quality of Service (QoS) and the transactional behavior are rather a measure on the whole message channel and can be determined through all involved EIPs. However, one characteristic that can be found in those contexts is the persistence requirement of a pattern (iii): all patterns that do not store whole messages or parts of them are \emph{stateless} pattern (\eg, typically all filters, message translators, routing). If a pattern requires to persist data to fulfill its purpose, it is \emph{stateful} (\eg, typically a \emph{Message Store}\index{Message Store (555)}, \emph{Aggregator}\index{Aggregator (268)}, \emph{Resequencer}\index{Resequencer (283)}). The characteristic of storing data in memory between invocations is not covered.

The very nature of a pattern is described by its specific processing logic (iv). This aspect of the integration semantics can be very specific to the underlying runtime implementation and is normally fully expressed in the available programming languages on the runtime platform.

The ability to follow a message from the sender to the receivers is very important. The information necessary for that is called lineage or message correlation (v). Patterns that change the original message or even generate new messages have to be able to link the (new) outgoing message(s) to the original message.

The processing types (vi) specify the runtime behavior of the pattern in terms of (a) sequential or parallel processing, (b) stream processing, (c) timeouts, and (d) behavior in exceptional situations. The capability to process messages in parallel (a) allows sending several messages to that pattern at a time (concurrently), while the sender still has to wait for the operation to be completed. When streaming is enabled, bigger data streams become tractable and the timeout (c) ensures that the processing stops after a configurable amount of time. If one of the patterns raises an exception during processing, its behavior shall be configurable (d) to either stop the processing of this pattern immediately or continue with other aspects of that pattern or even with their patterns.

\subsection{Integration Pattern Syntax and Model}
We now present the actual syntactic translation of the EIPs to BPMN as continuation of \cite{tuprints2751,volkersbuch} and specify the integration semantics in an informal manner by combining the work from \cite{Hohpe:2003:EIP:940308,DBLP:phd/de/Scheibler2010} with our experience. The subsequent descriptions are grouped by the different pattern categories and the numbers in brackets represent the pages for reference in \cite{Hohpe:2003:EIP:940308}. We start by describing a \emph{Message Channel} (60)\index{Message Channel (60)}, the fundamental infrastructure of a messaging system. The message channel connects sender with receiver applications (\ie, participants) for message-based communication. A simple form of a message channel is the \emph{Point-to-Point Channel}\index{Point-to-Point Channel (103)}, which connects one sender with one receiver directly. The unidirectional P2P channel is technically a bucket. The sender puts data on the channel (\ie, a message), the receiver takes it from the channel. If multiple receivers are connected to the channel, the actual recipient is not necessarily determined. However, a particular message is taken only by one receiver. The order of messages is a matter of implementation. In existing messaging systems we can normally assume the channel to be a queue, thus the messages are received in order of their sending.

Alternatively, one to many channels like publish-subscribe, broadcast, multi-cast and bus-like communication can be realized with message channels. A message sent to such a channel can be received by multiple receivers. For $n$ receivers, $n$ copies of the original message have to be provided.

In general, channels have special non-functional qualities like QoS characteristics that have to be configurable: \emph{Message Exchange Pattern} (MEP), \emph{availability}, \emph{security}, \emph{transactional processing}, \emph{Guaranteed Delivery} (122)\index{Guaranteed Delivery (122)}, and \emph{maximal message size}. The MEP defines a quality of the channel that can be derived from the adapter (\ie, message flow) or component/processor (\eg, activities, events) configurations. For instance, a file polling message flow acts one-way (InOnly), since it cannot handle response messages. On the other hand, most document message exchange works according to the \emph{Request-reply Pattern}\index{Request-Reply (154)}, which specifies a two way communication (InOut). In other words, channels can derive MEP constraints from its adapters and components or configure its behavior explicitly. A \emph{Channel} acts as logical address, thus the actual receiver is determined by the messaging system. A single \emph{Message}\index{Message (66)} can transport a piece of data (\ie, \emph{Document Message} (147)), a command for execution (\ie, \emph{Command Message} (145)\index{Command Message (145)}), or an event for logging (\ie, \emph{Event Message} (151)\index{Event Message (151)}). \emph{Pipes and Filters} \cite{DBLP:conf/ifip/Kahn74} as well as message routers are meant for influencing message content and direction. A filter can drop unwanted messages or content and a router can direct messages based on their content or a system's state to a particular destination, and a pipe as special form of message channel connects these components. A \emph{Message Translator}\index{Message Translator (85)} converts messages like transforming from one data format into another or extracting only necessary parts of a message. Finally, a \emph{Message Endpoint} (95)\index{Message Endpoint (95)} connects an application to a messaging system. In case of BPMN, the \emph{MessageFlow} represents a message endpoint by specifying the message with its structure, operation and interface (\eg, WSDL) that can be routed to the message channel.

\noindent
\textbf{Special Messaging Channels} \label{paragraph:smc}
Apart from the general channel specification there are some special channel types that are important. The \emph{Datatype Channel} (111)\index{Datatype Channel (111)} is a channel that only transports messages of a specific type. The message cardinality is 1:1, while the outgoing is the same as the incoming message. For exceptional cases, a channel for the storage and re-sending of messages can be configured (\eg, invalid message channel, dead letter channel). An \emph{Invalid Message Channel} (115)\index{Invalid Message Channel (115)} forwards messages, which could not be processed by the receiver. Its message cardinality is 1:0..1 depending on whether the outgoing message shall be stored or forwarded. In contrast, a \emph{Dead Letter Channel} (119)\index{Dead Letter Channel (119)} handles exceptional messages that have not yet been sent to a receiver, however, have been expired. The expiration is not determined by the dead letter channel, but by the base channel patterns (\eg, P2P, Publish/Subscribe). The message cardinality is the same as in the latter case.

\newcommand{\itembeg}[1]{\begin{minipage}[t]{#1}}
\newcommand{\itemend}{\end{minipage}}

\newcommand{\langfirst}{0.13\textwidth} 
\newcommand{\langsecon}{0.87\textwidth} 

\noindent
\textbf{Message Construction} When applications want to exchange data, they wrap it into a message. The message channels can only transport data wrapped in a message. The following patterns are subsequently discussed:
\begin{enumerate}
	\item Request-Reply (154)
	\begin{itemize}
		\item Synchronous Request-Reply\index{Request-Reply (154)}
		\item Asynchronous Request-Reply\index{Request-Reply (154)}
		\item Synch-Asynch Bridge\index{Synch-Asynch Bridge}
		\item Asynch-Synch Bridge\index{Asynch-Synch Bridge}
	\end{itemize}
	\item Correlation Identifier (163)
\end{enumerate}

\begin{table} \label{todo}
\caption{Request-Reply (154)\index{Request-Reply (154)|textbf}} \label{eip:requestreply}
\begin{tabular}{l*{2}{l}r}
\hline
Pattern              & Request-Reply (154)\index{Request-Reply (154)|textbf} \\
\hline
Model & \parbox[t]{0.9\columnwidth}{The \emph{Request-Reply} pattern is a communication pattern allowing a MEP with two-way (InOut) communication between two applications (\ie, in contrast to asynchronous, one-way communication). In particular, one application (requester) sends a request on one channel and waits for the response. The second application (replier) receives the request message and replies on a separate channel. The invocation of the requester can be synchronous (blocking) or asynchronous (non-blocking). The request channel can be a \emph{Point-to-Point Channel} (103)\index{Point-to-Point Channel (103)} or a \emph{Publish-Subscribe Channel} (105)\index{Publish-Subscribe Channel (105)}. The intend of the message exchange can be \emph{Messaging RPC} (RPC with a \emph{Command Message} request and a \emph{Document Message} response), \emph{Messaging Query} (remote query using a command message with the query and a document message with the result), or \emph{Notify/Acknowledge} (the request is an Event Message and the reply a Document Message with the acknowledgement). Hereby, the request can be either \emph{void} (\eg, notification), \emph{result value} (single result object), or \emph{Exception} (single exception object). The request can contain a return address and the reply should contain a \emph{Correlation Identifier}\index{Correlation Identifier (163)} to correlate to the request.} \\
Syntax & \parbox[t]{0.9\columnwidth}{
The request-reply stands for a synchronous or asynchronous communication with an external application (participant). Figure \ref{fig:requestreply} shows how a synchronous request is expressed by a \emph{Service Activity} that initiates the message processing in the external processing upon activation and data input (\ie, association to \emph{Data Object}). The activity waits until the external participant completed its processing and returns a message that is then assigned to the outgoing data object. In case of a \emph{fault} is returned by the service, it is treated as \emph{interrupting error} and the activity fails. The exception is processed in a separate channel.

An asynchronous request-reply is shown in Figure \ref{fig:requestreply_async}. The \emph{Send Activity} assigns the input from the \emph{Data Object} to the message and sends it to the external receiver. The activity is finished and a \emph{Receive Activity} is activated. The activity starts waiting for the associated message and the arriving message is assigned to the data object assigned to the activity. The correlation between the message sent to and the message received from the receiver have to be modeled in the definition of the \emph{Message} element of the \emph{Message Flow}.

Frequently used combinations of the previously discussed variants is the synch-asynch bridge (Figure \ref{fig:synch_asynch_bridge})\index{Synch-Asynch Bridge|textbf} and the asynch-synch bridge (Figure \ref{fig:asynch_synch_bridge})\index{Asynch-Synch Bridge|textbf}. While the asynch-synch bridge simply combines the shown syntax, the synch-asynch bridge in Figure \ref{fig:synch_asynch_bridge} shows an allowed syntactic variation by expressing the request by a \emph{Message Intermediate Event (throw)} that initiates the message processing in the external participant, while giving the control to a \emph{Message Intermediate Event (catch)} (\ie, activates the event) that waits for the response from the external participant. If and only if the participant responds with a message, the active event starts working. The syntax for the bridge patterns shown in Figures \ref{fig:synch_asynch_bridge} and \ref{fig:asynch_synch_bridge} requires an additional participant.} \\
\end{tabular}
\label{tab:cases}
\end{table}

\begin{figure}
\centering
\includegraphics[width=0.9\columnwidth]{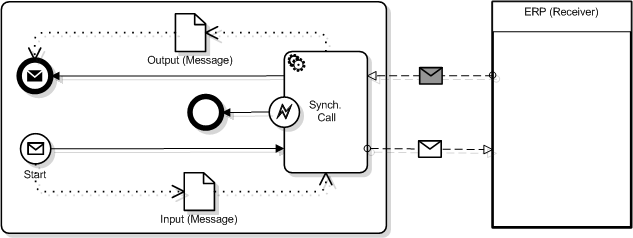}
\caption{Request-Reply Pattern (synchronous).}
\label{fig:requestreply}
\end{figure}

\begin{figure}
\centering
\includegraphics[width=0.9\columnwidth]{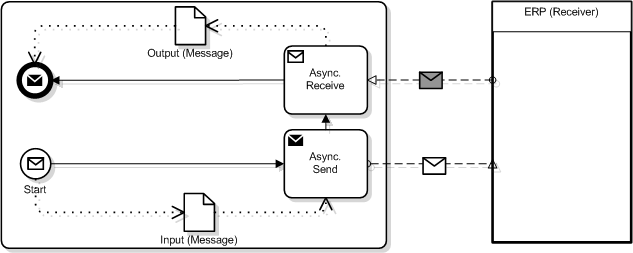}
\caption{Request-Reply Pattern (asynchronous).}
\label{fig:requestreply_async}
\end{figure}

\begin{figure}
\centering
\includegraphics[width=0.9\columnwidth]{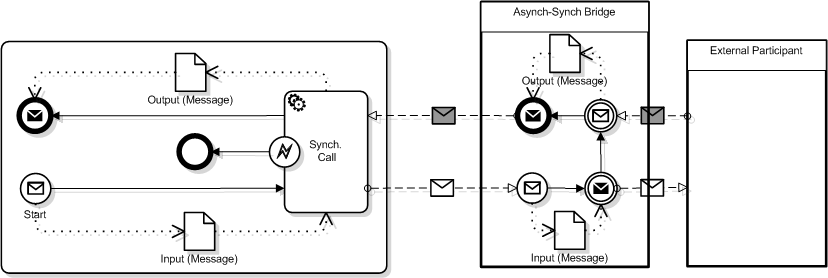}
\caption{Synch-Asynch Bridge.}
\label{fig:synch_asynch_bridge}
\end{figure}

\begin{figure}
\centering
\includegraphics[width=0.9\columnwidth]{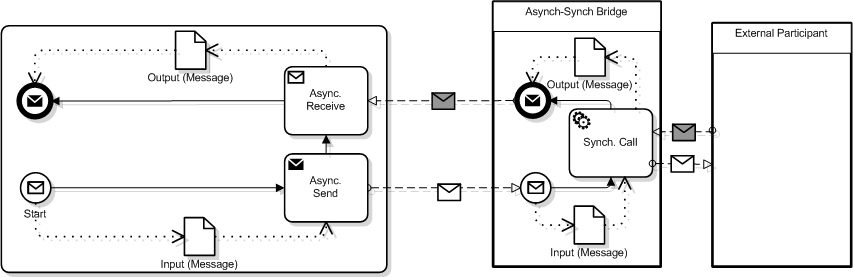}
\caption{Asynch-Synch Bridge.}
\label{fig:asynch_synch_bridge}
\end{figure}

\begin{table} \label{todo}
\caption{Correlation Identifier (163)\index{Correlation Identifier (163)|textbf} \label{eip:correlationid}} \label{todo}
\begin{tabular}{l*{2}{l}r}
\hline
Pattern              & Correlation Identifier (163)\index{Correlation Identifier (163)|textbf} \\
\hline
Model & \parbox[t]{0.9\columnwidth}{The \emph{Correlation Identifier} is used to correlate sent and received messages, when dealing with a multitude of messages. For instance, it can be used in a two-way message exchange for reply correlation to the request. The identifier mostly is a message header information.} \\
Syntax & \parbox[t]{0.9\columnwidth}{The correlation identifier is part of the message and the data objects that are part of the data flow. Hence there is no explicit syntax for it.} \\
\end{tabular}
\label{tab:cases}
\end{table}

\newpage

\noindent
\textbf{Message Routing} Messaging systems use message routing to decouple a message sender from the receiver of the message. In \cite{Hohpe:2003:EIP:940308} different types of routers are categorized: \emph{simple} (message routing from sender to one or more receivers), \emph{composed} (combined simple routers), and \emph{architectural patterns} (architectural styles based on routers). In this paper, we exclusively focus on simple routers and leave the task of composition up to the reader. The following patterns are subsequently discussed:
\begin{enumerate}
	\item Content-based Router (230)\index{Content-based Router (230)}
	\item Message Filter (237)\index{Message Filter (237)}
	\item Recipient List (249)\index{Recipient List (249)}
	\item Splitter (259)\index{Splitter (259)}
	\item Aggregator (268)\index{Aggregator (268)}
	\item Resequencer (283)\index{Resequencer (283)}
\end{enumerate}

\begin{table} \label{todo}
\caption{Content-based Router (230)\index{Content-based Router (230)|textbf}} \label{todo}
\begin{tabular}{l*{2}{l}r}
\hline
Pattern              & Content-based Router (230)\index{Content-based Router (230)|textbf} \\
\hline
Model & \parbox[t]{0.9\columnwidth}{The \emph{Content-based Router} determines the receiver of an incoming message based on its content. Dynamically forwarding the message to the correct receiver requires knowledge about all possible recipients. The router receives exactly one incoming message and routes it to one of multiple outbound channels. Neither the message nor its content is changed. The router is stateless.} \\
Syntax & \parbox[t]{0.9\columnwidth}{The content-based router decides on the route of the messages represented by a \emph{Data Object} by evaluating a \emph{Condition Expression} on the \emph{Sequence Flow}. The sequence flow syntax reads as follows: if no condition (diamond shape) evaluates true, the default sequence flow is taken (diagonal dash; Figure \ref{fig:contentbasedrouter}). Since the pattern does not change the input message, the output data objects could be neglected (syntactically) during modeling.

\centering
\includegraphics[width=0.7\columnwidth]{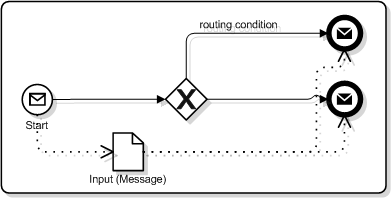}
\caption{Content-based Router Pattern.}
\label{fig:contentbasedrouter}} \\
Summary & \parbox[t]{0.9\columnwidth}{1:n message cardinality, 1:n channel cardinality (same outbound message formats), non-message generating, stateless, routing condition.} \\
\end{tabular}
\label{tab:cases}
\end{table}

\begin{table} \label{todo}
\caption{Message Filter (237)\index{Message Filter (237)|textbf}} \label{todo}
\begin{tabular}{l*{2}{l}r}
\hline
Pattern              & Message Filter (237)\index{Message Filter (237)|textbf} \\
\hline
Model & \parbox[t]{0.9\columnwidth}{The \emph{Message Filter} is a routing pattern with exactly one outbound channel. In case the content of the incoming message matches the criteria specified in the filter condition, the message is routed to the output channel. Else the message is discarded. The filter is stateless.} \\
Syntax & \parbox[t]{0.9\columnwidth}{The message filter receives one input message as \emph{Data Object} and uses a static filter condition (again specified as \emph{Script Task}) to emit only those messages sufficient to the condition. Figure \ref{fig:message_filter} shows the pattern with a default message channel and a conditional one that does not pass the filtered messages further.

\centering
\includegraphics[width=0.7\columnwidth]{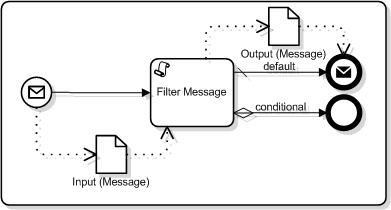}
\caption{Message Filter Pattern.}
\label{fig:message_filter}} \\
Summary & \parbox[t]{0.9\columnwidth}{1:1 message cardinality, 1:1 channel cardinality, non-message generating, stateless, filter expression.} \\
\end{tabular}
\label{tab:cases}
\end{table}

\begin{table} \label{todo}
\caption{Recipient List (249)\index{Recipient List (249)|textbf}} \label{todo}
\begin{tabular}{l*{2}{l}r}
\hline
Pattern              & Recipient List (249)\index{Recipient List (249)|textbf} \\
\hline
Model & \parbox[t]{0.9\columnwidth}{The \emph{Recipient List} emits an incoming message to several receivers by copying in the original message. In contrast to the \emph{Publish/Subscribe}\index{Publish-Subscribe Channel (105)} pattern the recipient list controls the determination of the receivers. The recipient list consists of two configurable parts: the receiver determination and the distribution of the message copies to the receivers, while the message remains unchanged. The receiver determination can be statically defined, contained in the message or computed from the message. If the list of recipients is part of the message, it can be removed from the message to hide it from the receivers, which would be done by adding a \emph{Message Translator}\index{Message Translator (85)} before the sending step. The recipient list processes outbound messages in order of the output channels (\ie, no parallel sending). With that characteristic, prioritized outbound processing can be modeled.

For robustness reasons (\eg, transactional processing), the recipient list can be stateful (default: stateless). The pattern has to ensure that the incoming message is successfully delivered to all recipients. In other words this operation has to be atomic and restartable to complete all incomplete operations. To achieve that, the recipient list (a) can use transactional outbound channels on which the messages are put successfully before committing them, which guarantees an ``all or nothing" behavior (\emph{Single Transaction}), (b) can store the information about the sending status for the restart of incomplete messages after an incident (\emph{Persistent recipient list}), or (c) can resend all messages--independent of their status, which requires all recipients to be \emph{Idempotent Receivers}. In case of (c), the \emph{Message Filter}\index{Message Filter (237)} pattern can be used to eliminate duplicates, which would make a receiver idempotent. For (b) the recipient list must be configured stateful.} \\
Syntax & \parbox[t]{0.9\columnwidth}{The recipient list emits copies of incoming messages (\emph{Data Object}) to a list of statically or dynamically specified recipients. Figure \ref{fig:recipientlist} the pattern with a data object representing the static list of recipients and several conditional receivers.

\centering
\includegraphics[width=0.7\columnwidth]{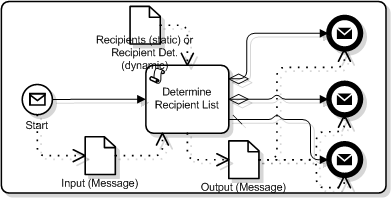}
\caption{Recipient List Pattern.}
\label{fig:recipientlist}} \\
Summary & \parbox[t]{0.9\columnwidth}{1:n message cardinality, 1:n channel cardinality, message generating (copy), stateless (or stateful for robustness reasons), recipient list (incl. mapping to receivers) or receiver determination program, ordered processing.} \\
\end{tabular}
\label{tab:cases}
\end{table}

\begin{table} \label{todo}
\caption{Splitter (259)\index{Splitter (259)|textbf} \label{eip:splitter}} \label{todo}
\begin{tabular}{l*{2}{l}r}
\hline
Pattern              & Splitter (259)\index{Splitter (259)|textbf} \label{eip:splitter} \\
\hline
Model & \parbox[t]{0.9\columnwidth}{A \emph{Splitter} breaks one original message into multiple (smaller) messages. For 
that, the splitter creates as many new messages as the split results to. Common repeated elements (\eg, 
correlation identifiers, body elements) are duplicated and added to each new message. Per default, the 
resulting messages are ordered by the sequence they are processed.  In general, the splitter can emit 
messages ordered or unordered. The default ordering is determined by the sequence of split operations. A 
sequence number can be added for later resequencing (\ie, Resequencer Pattern (283)\index{Resequencer (283)} or aggregation (\ie, Aggregator Pattern (268)\index{Aggregator (268)}). Adding 
the original message id to the produced messages as Correlation Identifier (163)\index{Correlation Identifier (163)} could help tracing messages and is useful for Request-Reply pattern (154)\index{Request-Reply (154)}.

A splitter that is able to split tree structures into sub-trees and for which the number of generated 
messages is unknown, but those messages always have the same structure, is called \emph{Iterative Splitter} 
(Figure \ref{fig:splitter}). In contrast, a \emph{Static Splitter} splits a message structure into a fixed 
number of parts, while knowing the schema. The structure of the generated messages can differ (Figure \ref{fig:splitter3}).

Originally, the splitter is a stateless pattern (i.e, the messages are not stored during processing). The 
splitter could be modeled as multi-cast channel followed by a set of content filters on the outgoing 
channels. In the case of a static splitter, the message structure of the outgoing messages can 
vary.} \\
Syntax & \parbox[t]{0.9\columnwidth}{The static and the iterative splitters break a message into smaller messages according 
to a split expression either provided as additional \emph{Data Object} (not shown) or by using a \emph{Script Task}. The script is platform/ engine specific and is executed when the task is performed. The structure of the input and output messages are 
again defined by a data object. If the message format is the same for all outgoing messages, the data object 
representing the output messages is configured as \emph{Data Object Collection} (Figure \ref{fig:splitter}). In contrast to \cite {DBLP:phd/de/Scheibler2010}, the number of output channels is assumed to be one, thus it cannot be parameterized. However, \emph{Message End Events} in BPMN cannot emit messages of different type. In case of different message formats, a ``Parallel Gateway" is used to emit the messages to different end events depending on its format. Therefore in our syntax, several end events and thus several outbound channels are addressable.} \\
Summary & \parbox[t]{0.9\columnwidth}{1:n message cardinality, 1:1 channel cardinality (1:n in case of differing outbound message formats), message generating, stateless, correlation identifier, 
split expression.} \\
\end{tabular}
\label{tab:cases}
\end{table}

\begin{figure}
\centering
\includegraphics[width=0.9\columnwidth]{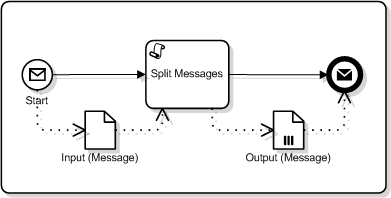}
\caption{Message Splitter Pattern.}
\label{fig:splitter}
\end{figure}

\begin{figure}
\centering
\includegraphics[width=0.9\columnwidth]{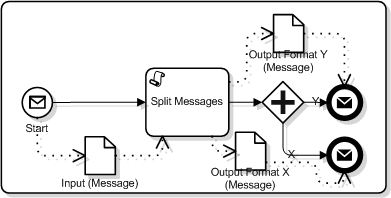}
\caption{Message Splitter Pattern with differing output messages.}
\label{fig:splitter3}
\end{figure}

\begin{table} \label{todo}
\caption{Aggregator (268)\index{Aggregator (268)|textbf} \label{eip:aggregator}} \label{todo}
\begin{tabular}{l*{2}{l}r}
\hline
Pattern              & Aggregator (268)\index{Aggregator (268)|textbf} \label{eip:aggregator} \\
\hline
Model & \parbox[t]{0.9\columnwidth}{An \emph{Aggregator} combines single, but related, incoming messages to a new, complex message. For that, it receives a stream of messages and identifies correlated messages, until a complete set of correlated messages has received. The aggregator applies an aggregation algorithm and emits a single message. An aggregator is a stateful pattern (\eg, for checking the completeness condition), however, not all messages have to be stored completely, but only information relevant for the aggregation. The aggregator consists of at least the following parts: (i) the aggregation algorithm, (ii) the completeness condition, (iii) the correlation function with Correlation Identifier (163)\index{Correlation Identifier (163)} in one or multiple fields in the body and header of the messages, and (iv) a aging/ removal strategy for the stored information.

For the aggration (i), the aggragator holds a list of aggregates. New messages/ information is either added to existing aggregates or create new ones (no prior knowledge required: self-starting aggregator). If messages arrive that belong to an already closed aggregate and the opening of new aggregates shall be prevented, the closed aggregates have to be stored. The closed aggregates list should be purged periodically (should not grow infinitely). For instance, Message Expiration (176) \cite{tuprints2751}\index{Message Expiration (176)} could be used for that. For the purge algorithm, not the complete messages have to be remembered, but only the close information. A Control Bus (540)\index{Control Bus (540)},
 could help to allow manual purging. The following types of aggregation algorithms have to be considered: select the best answer, condense data, and collect data for later evaluation (\ie, simply store data if the aggregator cannot make a decision). For the completeness condition (ii), the following \emph{Aggregation Strategies} have to be considered: wait for all, timeout, first best, timeout with override (break faster and override), external event (\eg, Control Bus\index{Control Bus (540)}).

A \emph{Static Aggregator} always aggregates a fixed number of messages. In contrast to \cite{DBLP:conf/caise/FahlandG13,tuprints2751}, only one incoming message channel is considered. The aggregator is the inverse of the Splitter Pattern (259)\index{Splitter (259)}} \\
Syntax & \parbox[t]{0.9\columnwidth}{The high-level syntax of the aggregator pattern is shown in Figure \ref{fig:aggregator_compact}. The first message that arrives instanciates an aggregator \emph{Sub-Process} as \emph{Activity} (collapsed). All subsequent messages are send to that sub-process directly. An \emph{Escalation Intermediate Event} is used to send one aggregate to the receiver. The aggregation can timeout through a \emph{Timer Intermediate Event}, which stops the execution.

The expanded activity is shown in Figure \ref{fig:aggregator}. The ``loop-wait" nature of the pattern is indicated by the ``cyclic arrow" on the bottom of the sub-process. The structures of the incoming messages and the outgoing (aggregated) messages are specified via a \emph{Data Object}. The correlation function and the correlation identifier (iii) to correlate different messages to one are defined by an \emph{Activity} in the \emph{Sub-Process} definition. The completeness conditions (ii) are represented by an associated \emph{EventDefintion} of type \emph{Interrupting Escalation StartEvent} (\eg, for wait for all, first best strategies) or of type \emph{Interrupting Timer StartEvent} (\eg, timeout (with override), completion interval). The aggregation algorithm (i) is again denoted by an activity that specifies the named algorithms (\eg, a greater-equals operation on a special element of the message payload for a best-answer algorithm). In addition, a message translation algorithm can be specified to combine the incoming messages to one has to be provided. Open topics: external event and connection to the control bus.} \\
Summary & \parbox[t]{0.9\columnwidth}{n:1 messages (not channels), message generating (combines several original messages to 
one new message), stateful, aggregation algorithm, completeness condition, correlation function, 
correlation identifier, massege translation program.} \\
\end{tabular}
\label{tab:cases}
\end{table}

\begin{figure}
\centering
\includegraphics[width=0.9\columnwidth]{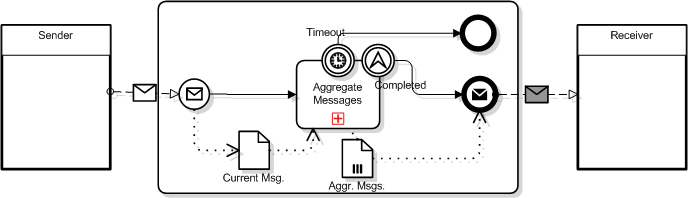}
\caption{Aggregator Pattern (collapsed).}
\label{fig:aggregator_compact}
\end{figure}

\begin{figure}
\centering
\includegraphics[width=0.9\columnwidth]{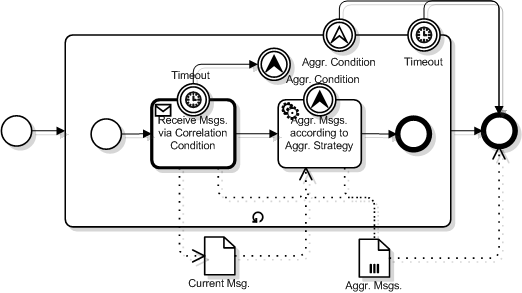}
\caption{Aggregator Pattern (expanded).}
\label{fig:aggregator}
\end{figure}

\begin{table} \label{todo}
\caption{Resequencer (283)\index{Resequencer (283)|textbf} \label{eip:resequencer}} \label{todo}
\begin{tabular}{l*{2}{l}r}
\hline
Pattern              & Resequencer (283)\index{Resequencer (283)|textbf} \label{eip:resequencer} \\
\hline
Model & \parbox[t]{0.9\columnwidth}{The \emph{Resequencer} orders a stream of un-ordered messages according to a sequence number and emits the messages to the output channel in the proper sequence. The concept of a unique sequence number has to be attached to each message. Hence the in-order sequence numbers distinguish from \emph{Correlation Identifiers} and message identifiers. To handle cases, in which messages arrive unordered, the resequencer has to be stateful. These stored messages have to be checked for the completeness of the sequence to decide on when to order and emit the messages. When the ordered set of messages was successfully sent, the stored messages have to be removed from the resequencer's message store.} \\
Syntax & \parbox[t]{0.9\columnwidth}{The resequencer takes a set of messages as \emph{Data Object Collection} and orders them according to a specified sequence that is emitted represented by a \emph{Data Object}. Figure \ref{fig:resequencer} shows the resequencer as collapsed \emph{Sub-Process}. When expanding the sub-process (Figure \ref{fig:resequencer_exp}), an activated \emph{Activity} receives the messages and stores them in a store (\emph{Service Activity} associated to a \emph{Data Object}). When the complete sequence has been received, another activated activity takes the list of messages and emits them in order. In contrast to the other patterns, the resequencer in BPMN has an impact on the complete syntax in one participant. If the integration logic does not only re-orders messages, other steps have to be added as pre-or post processing to Figure \ref{fig:resequencer_exp}.

%
} \\
Summary & \parbox[t]{0.9\columnwidth}{1:1 messages (different order), non-message generating, stateful, sequence number, sequence complete condition, order operation.} \\
\end{tabular}
\label{tab:cases}
\end{table}

\begin{figure}
\centering
\includegraphics[width=0.9\columnwidth]{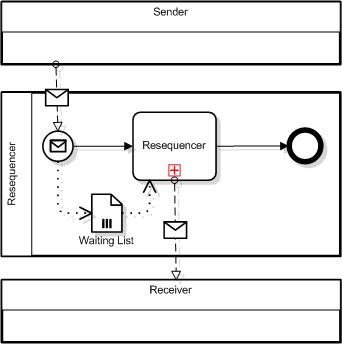}
\caption{Resequencer Pattern (collapsed).}
\label{fig:resequencer}
\end{figure}

\begin{figure}
\centering
\includegraphics[width=0.9\columnwidth]{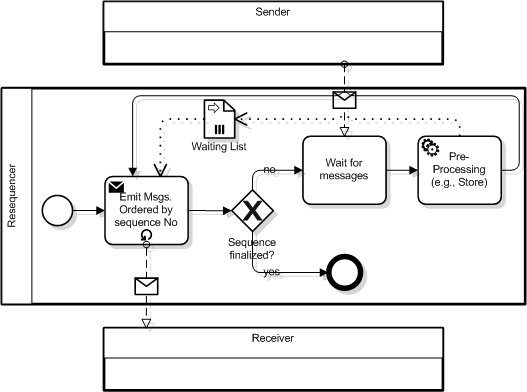}
\caption{Resequencer Pattern (expanded).}
\label{fig:resequencer_exp}
\end{figure}

\newpage

\noindent
\textbf{Message Transformation} Since sender and receiver applications rarely agree on a common data format, messages have to be changed within the messaging (channel). The following patterns are subsequently discussed:
\begin{enumerate}
	\item Content Enricher (336)\index{Content Enricher (336)}
	\item Content Filter (342)\index{Content Filter (342)}
	\item Claim Check (346)\index{Claim Check (346)}
\end{enumerate}

\begin{table}[h] \label{todo}
\caption{Message Translator (85)\index{Message Translator (85)|textbf} \label{eip:translator}} \label{todo}
\begin{tabular}{l*{2}{l}r}
\hline
Pattern              & Message Translator (85)\index{Message Translator (85)|textbf} \label{eip:translator} \\
\hline
Model & \parbox[t]{0.9\columnwidth}{A \emph{Message Translator} converts an incoming message (format) into a data format expected by the corresponding receiver of the message. The concrete transformation logic is represented by a mapping definition that allows different levels of transformations, at least: (a) data structure (\eg, n:m structure mappings like projections), (b) data (\eg, value mapping), (c) data types (\eg, date/time conversion), (d) data representation (\eg, encryption, compression, data formats like XML, JSON), and (e) the transport protocol (\eg, TCP/IP). The translator is stateless and does modify the original message.} \\
Syntax & \parbox[t]{0.9\columnwidth}{The incoming message (format), outgoing message (format) as well as the mapping logic are specified by a \emph{Data Object} (Figure \ref{fig:messagetranslator}).

\centering
\includegraphics[width=0.9\columnwidth]{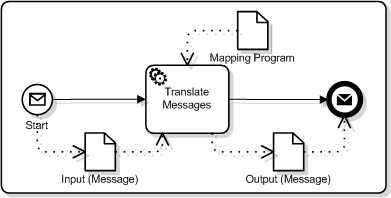}
\caption{Message Translator Pattern.}
\label{fig:messagetranslator}} \\
Summary & \parbox[t]{0.9\columnwidth}{1:1 message cardinality, non-message generating, stateless, mapping program.} \\
\end{tabular}
\label{tab:cases}
\end{table}

\begin{table} \label{todo}
\caption{Content Enricher (336)\index{Content Enricher (336)|textbf}} \label{todo}
\begin{tabular}{l*{2}{l}r}
\hline
Pattern              & Content Enricher (336)\index{Content Enricher (336)|textbf} \\
\hline
Model & \parbox[t]{0.9\columnwidth}{The \emph{Content Enricher} allows to add information to an incoming message. The additional content can be assigned locally, \ie, determined within the pattern, or can be requested from external participants. Hereby, the locally assigned information can come from a computation on the current message or from the runtime environment. The content enricher is stateless and generates a new message from the aggregation of the input message and the additional information. The message cardinality is 1:1, while the outgoing message might have a different format than the incoming message.} \\
Syntax & \parbox[t]{0.9\columnwidth}{The content enricher uses the synchronous \emph{Request-Reply}\index{Request-Reply (154)} pattern (\ie, using a \emph{Service Task}) to fetch data from an external participant and merges it with the incoming message represented as \emph{Data Object} (Figure \ref{fig:content_enricher}). Hereby, the service task specifies the operation that is used to create the request (incl. mapping) and the assigned \emph{Message Flow} specifies the (remote) endpoint (\eg, transport-, message protocol).

\begin{center}
\includegraphics[width=0.9\columnwidth]{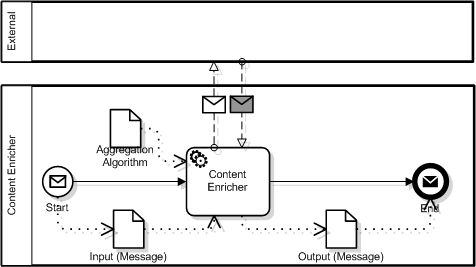}
\caption{Content Enricher Pattern.}
\label{fig:content_enricher}
\end{center}} \\
Summary & \parbox[t]{0.9\columnwidth}{1:1 message cardinality (format might change) and message channels within the process and 1:1 message flow channel to the external service, message generating (passes response), stateless, request mapping program, response aggregation/ enrichment.} \\
\end{tabular}
\label{tab:cases}
\end{table}

\begin{table} \label{todo}
\caption{Content Filter (342)\index{Content Filter (342)|textbf}} \label{todo}
\begin{tabular}{l*{2}{l}r}
\hline
Pattern              & Content Filter (342)\index{Content Filter (342)|textbf} \\
\hline
Model & \parbox[t]{0.9\columnwidth}{The \emph{Content Filter} removes parts of the message not required by the recipient, 
or it simplifies the data structure through, \eg, thinning, flattening complex structures via projection 
according to its filter condition. Alternatively, the filter condition can come from an \emph{External Service}\index{External Service}. A filter is a stateless pattern, which works on the original (incoming) message (\ie, does not generate new messages). A content filter is a special variant of a Message 
Translator\index{Message Translator (85)}, while not translating the whole message, but only the 
required parts. A sequence of content filters can be used as static splitters.} \\
Syntax & \parbox[t]{0.9\columnwidth}{The content filter (shown in Figure \ref{fig:content_filter}) specifies a filter 
condition, the structure of the incoming and the outgoing message as \emph{Data Object}.

\centering
\includegraphics[width=0.9\columnwidth]{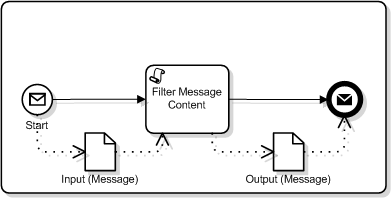}
\caption{Content Filter Pattern.}
\label{fig:content_filter}} \\
Summary & \parbox[t]{0.9\columnwidth}{1:1 messages, non-message generating, stateless, filter condition.} \\
\end{tabular}
\label{tab:cases}
\end{table}

\begin{table} \label{todo}
\caption{Claim Check (346)\index{Claim Check (346)|textbf}} \label{todo}
\begin{tabular}{l*{2}{l}r}
\hline
Pattern              & Claim Check (346)\index{Claim Check (346)|textbf} \\
\hline
Model & \parbox[t]{0.9\columnwidth}{The \emph{Claim Check} allows to model a ``call by reference" for the complete payload of a message or only parts of it. The data from the message payload is removed temporarily from the message (\eg, if it is not relevant for the next-n processing steps) and can be added to the message later. For that, a filter condition is used to extract the data from the message and a unique key is generated under which the data is stored in a data store. The key is added to the message (\eg, as header field) and the message is sent to the channel. Later the key is used to ``claim" the data from the data store and a local \emph{Content Enricher} is used to add the data to the message. The content enricher has to take care that the ``claim check" key is removed from the message afterwards. This pattern can be seen as the counterpart to a \emph{Content Enricher}\index{Content Enricher (336)}.} \\
Syntax & \parbox[t]{0.9\columnwidth}{The claim check frees an incoming message (\emph{Data Object}) from its payload by storing the payload in a \emph{Data Store}, while only passing the message without its payload and a claim check as data object to the next flow element (Figure \ref{fig:claim_check}). To retrieve the payload, a local \emph{Content Enricher}\index{Content Enricher (336)} as \emph{Script Task} (from store; normally \emph{Service Activity}) gets the claims check and the corresponding payload from the data store.

\centering
\includegraphics[width=0.9\columnwidth]{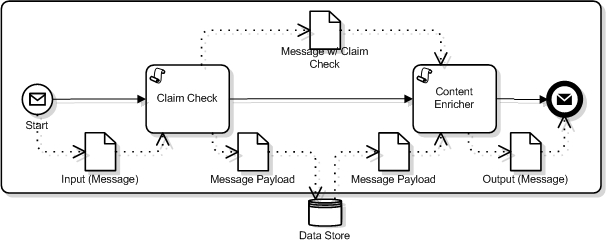}
\caption{Claim Check Pattern.}
\label{fig:claim_check}} \\
Summary & \parbox[t]{0.9\columnwidth}{1:1 messages, non-message generating, stateful, filter condition, local \emph{Content Enricher}\index{Content Enricher (336)} (from store).} \\
\end{tabular}
\label{tab:cases}
\end{table}

\newpage

\noindent
\textbf{System Management} The operation of a messaging system is equally challenging as its implementation itself. For that, some patterns have been compiled by \cite{Hohpe:2003:EIP:940308}, which are discussed subsequently. The following patterns are subsequently discussed:
\begin{enumerate}
	\item Wire Tap (547)\index{Wire Tap (547)}
	\item Message Store (555)\index{Message Store (555)}
\end{enumerate}




\begin{table} \label{todo}
\caption{Wire Tap (547)\index{Wire Tap (547)|textbf}} \label{todo}
\begin{tabular}{l*{2}{l}r}
\hline
Pattern              & Wire Tap (547)\index{Wire Tap (547)|textbf} \\
\hline
Model & \parbox[t]{0.9\columnwidth}{The \emph{Wire Tap} allows access to messages in a point-to-point channel. The wire tap is comparable to a static recipient list with two recipients and two outgoing message channels. While the first channel is the normal message channel, the second channel transports a copy of the original message for analysis or storage. Typically the extra channel is a point-to-point channel that persists the message in a \emph{Message Store}\index{Message Store (555)}.} \\
Syntax & \parbox[t]{0.9\columnwidth}{Figure \ref{fig:wiretap} shows the usage of a wire tap with a point-to-point channel that persists the message in a message store. The input, output and copy of output messages are specified by a \emph{Data Object}. The static routing is defined by an \emph{Parallel Gateway}, which is a variant to the syntax shown for the recipient list pattern. In the example, the channel the data flow ends with the storage of the message.

\centering
\includegraphics[width=0.9\columnwidth]{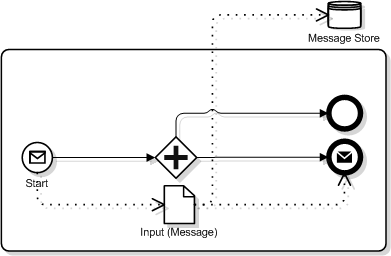}
\caption{Wire Tap Pattern.}
\label{fig:wiretap}} \\
Summary & \parbox[t]{0.9\columnwidth}{1:1 message cardinality, 1:2 message channels, message generating (copy), stateless.} \\
\end{tabular}
\label{tab:cases}
\end{table}

\begin{table} \label{todo}
\caption{Message Store (555)\index{Message Store (555)}} \label{todo}
\begin{tabular}{l*{2}{l}r}
\hline
Pattern              & Message Store (555)\index{Message Store (555)} \\
\hline
Model & \parbox[t]{0.9\columnwidth}{The \emph{Message Store} allows operations on messages like persist (insert), update, delete and query. The message store can be any kind of (No)SQL, in-memory or file-based storage for arbitrary message storage formats (\eg, BLOB, XML, JSON). Many persistent pattern (\eg, \emph{Aggregator}\index{Aggregator (268)}, idempotent receiver) use the store to guarantee the correct integration semantics. Other non-persistent patterns (\eg, wire tap) or integration semantics (\eg, \emph{Guaranteed Delivery}\index{Guaranteed Delivery (122)}, retry handling on channel level) can be assigned.} \\
Syntax & \parbox[t]{0.9\columnwidth}{Following the process model definition, associated \emph{Data Objects} define the flow of data through a message channel. The same mechanism is used to assign the message store (\ie, represented as \emph{Data Store}) to elements within a \emph{Process} via \emph{Data Association}. Since data stores are defined beyond the scope of a process in BPMN, the store can be associated by arbitrary many elements from different processes and can thus denote local or shared storages.} \\
\end{tabular}
\label{tab:cases}
\end{table}

\newpage

\noindent
\textbf{Additional Patterns} Not all patterns in scope of this document have been covered by \cite{Hohpe:2003:EIP:940308}. However, these patterns have been proposed and were partially taken up for implementation by \cite{DBLP:phd/de/Scheibler2010,apacheCamel13}. The following patterns are subsequently discussed:
\begin{enumerate}
	\item Join Router 
	\item External Service \index{External Service}
	\item Multicast \index{Multicast}
\end{enumerate}

\begin{table} \label{tab:joinrouter}
\caption{Join Router\index{Join Router|textbf}} \label{todo}
\begin{tabular}{l*{2}{l}r}
\hline
Pattern              & Join Router\index{Join Router|textbf} \\
\hline
Model & \parbox[t]{0.9\columnwidth}{The \emph{Join Router} is a structuring element that combines several different incoming channels to one outgoing channe. Received messages from different channels are passed while not executing any operation on them. The join router is stateless. It is assumed that the formats of the incoming messages has to be the same (\eg, enforced by pre-processed \emph{Message Translators}\index{Message Translator (85)}).} \\
Syntax & \parbox[t]{0.9\columnwidth}{The join route combines a several channels to one without specifying a merge condition. Figure \ref{fig:joinrouter} shows the usage of an \emph{Inclusive Gateway} that is activated if at least one incoming flow passes the control to the gateway and emits a message if an incoming message via \emph{Data Object} is received. In other words, all incoming messages are simply routed to one channel.

\centering
\includegraphics[width=0.9\columnwidth]{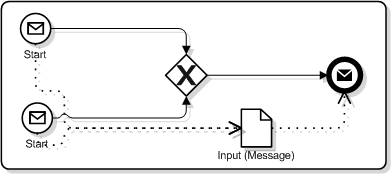}
\caption{Join Router Pattern.}
\label{fig:joinrouter}} \\
Summary & \parbox[t]{0.9\columnwidth}{1:1 message cardinality, n:1 message channels, non-message generating, stateless.} \\
\end{tabular}
\label{tab:cases}
\end{table}

\begin{table} \label{tab:externalservice}
\caption{External Service\index{External Service|textbf} \label{eip:externalservice}}
\begin{tabular}{l*{2}{l}r}
\hline
Pattern              & External Service\index{External Service|textbf} \label{eip:externalservice} \\
\hline
Model & \parbox[t]{0.9\columnwidth}{The \emph{External Service} is a building block for all patterns that require external data (\eg, \emph{Synch-Asynch Bridge}\index{Synch-Asynch Bridge}, \emph{Asynch-Synch Bridge}\index{Asynch-Synch Bridge}, \emph{Content Enricher}\index{Content Enricher (336)}). The pattern is stateless and has an internal and external message and channel cardinality of 1:1, respectively.} \\
Syntax & \parbox[t]{0.9\columnwidth}{Figure \ref{fig:externalservice} shows an external service that receives request messages defined through a \emph{Data Object} and emits the response message in the same way. The request is passed via \emph{Message Flow} to the external endpoint and the response from the endpoint is set on the channel for subsequent processing.

\centering
\includegraphics[width=0.9\columnwidth]{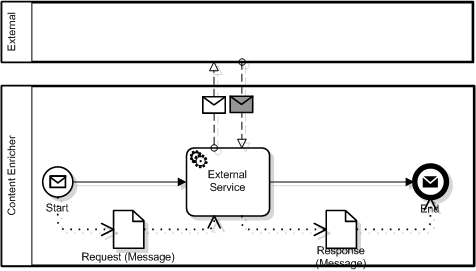}
\caption{External Service Pattern.}
\label{fig:externalservice}} \\
Summary & \parbox[t]{0.9\columnwidth}{1:1 message cardinality (format might change) and message channels within the process and 1:1 message flow channel to the external service, message generating (passes response), stateless.} \\
\end{tabular}
\label{tab:cases}
\end{table}

\begin{table} \label{tab:multicast}
\caption{Multicast\index{Multicast|textbf} \label{eip:multicast}}
\begin{tabular}{l*{2}{l}r}
\hline
Pattern              & Multicast\index{Multicast|textbf} \label{eip:multicast} \\
\hline
Model & \parbox[t]{0.9\columnwidth}{The \emph{Mutlicast} routes one incoming message to several receivers in parallel. For that, the incoming message is copied to multiple output channels. In contrast to the \emph{Recipient List}\index{Recipient List (249)} the receiver determination has to be (statically) configured and is not determined based on the incoming messages. The multi-cast is stateless and will continue processing messages to the separate output channels even if one of them fails. To reach the same robust messaging capabilities as the recipient list, the same configurations can be done (see Section \ref{sec:integrationsemantics} (a)--(c)).} \\
Syntax & \parbox[t]{0.9\columnwidth}{The multi-cast takes one incoming messages as \emph{Data Object} and emits several copies of the original message to the recipients via data objects. Figure \ref{fig:multicast} shows the syntax of the multi-cast using a \emph{Parallel Gateway}.} \\

 & \parbox[t]{0.9\columnwidth}{

\centering
\includegraphics[width=0.9\columnwidth]{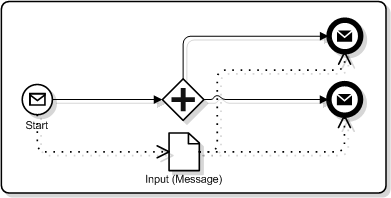}
\caption{Multicast Pattern.}
\label{fig:multicast}
} \\

Summary & \parbox[t]{0.9\columnwidth}{1:n message cardinality, 1:n message channels, message generating (copy message with new message identifier), stateless (similar to \emph{Recipient List}), behavior in exceptional cases (\eg, stop processing).} \\
\end{tabular}
\label{tab:cases}
\end{table}

\section{Concluding Remarks} \label{sec:conclusion}
The Enterprise Integration Patterns are a set of widely used patterns denoting the building blocks for a structured implementation of a messaging system. For these patterns we have continued the work of \cite{tuprints2751,volkersbuch} by specifying a syntactic mapping to the Business Process Model and Notation (BPMN); thus each pattern is a set of elements in a BPMN Process, which can be composed to sets of message channels from the senders to their receivers. Together with the syntax, we defined a process model for its semantics (referred to as model). Although the syntax is compliant to BPMN, the BPMN execution semantics are replaced by a process model, we slightly adapted to the integration domain. The result is an integration domain specific language, with which integration aspects of messaging systems and their runtimes can be expressed.


\section*{Acknowledgements}
Thanks to Volker Stiehl and Ivana Trickovic for valuable discussions on BPMN and the proposed syntax, Christian Mathis for the joint work on the implementation of a runtime system, Christian Becker and Markus M\"unkel for proof reading and valuable discussions.

\bibliographystyle{abbrv}
\bibliography{bpmn}

\newpage
\renewcommand{\indexname}{List of Patterns}
\printindex

\end{document}